\documentclass[aps,prl,twocolumn,preprintnumbers,superscriptaddress,groupedaddress,nofootinbib]{revtex4}
\pdfoutput=1

\usepackage{graphicx}     

\usepackage[bookmarksopen,colorlinks=true,linkcolor=steelblue,
citecolor=orangered,urlcolor=darkred,linktoc=all]{hyperref}

\usepackage{amsmath}
\usepackage{amsfonts}
\usepackage{amssymb}
\usepackage{bm}        
\usepackage{slashed}   

\usepackage{amsopn}
\mathchardef\mhyphen="2D

\usepackage{xcolor}
\definecolor{darkred}{rgb}{0.7, 0., 0.}
\definecolor{orangered}{rgb}{1,0.27,0.}
\definecolor{steelblue}{rgb}{0.275,0.51, 0.706}
\definecolor{forestgreen}{rgb}{0.13,0.55,0.13}

\begin{document}

\preprint{UMN-TH-4115/22,~~FTPI-MINN-22-06}

\title{\Large Standard Model prediction for paramagnetic EDMs}

\author{Yohei Ema}
\email{ema00001@umn.edu}
\affiliation{William I. Fine Theoretical Physics Institute, School of Physics and Astronomy,
University of Minnesota, Minneapolis, MN 55455, USA}
\author{Ting Gao}
\email{gao00212@umn.edu}
\affiliation{School of Physics and Astronomy, University of Minnesota, Minneapolis, MN 55455, USA}
\author{Maxim Pospelov}
\email{pospelov@umn.edu}
\affiliation{William I. Fine Theoretical Physics Institute, School of Physics and Astronomy,
University of Minnesota, Minneapolis, MN 55455, USA}
\affiliation{School of Physics and Astronomy, University of Minnesota, Minneapolis, MN 55455, USA}

\date{\today}

\begin{abstract}
Standard Model $CP$ violation associated with the phase of the Cabibbo-Kobayashi-Maskawa quark mixing matrix is known to give 
small answers for the EDM observables. Moreover, predictions for the EDMs of neutrons and diamagnetic atoms suffer from considerable uncertainties. We point out that the $CP$-violating observables associated with the electron spin (paramagnetic EDMs) are dominated by the combination of the electroweak penguin diagrams and $\Delta I =1/2$ weak transitions in the baryon sector, and are calculable within chiral perturbation theory. The predicted size of the semileptonic operator $C_S$ is $7\times 10^{-16}$ which corresponds to the {\em equivalent} electron EDM $d_e^{\rm eq} = 1.0 \times 10^{-35}e\,{\rm cm}$. While still far from the current observational limits, this result is three orders of magnitude larger than previously believed. 
\end{abstract}


\maketitle

\textbf{Introduction}\,---\,
The searches for EDMs of elementary particles~\cite{Graner:2016ses,Cairncross:2017fip,Andreev:2018ayy,nEDM:2020crw} represent an important way of probing the TeV scale new physics \cite{Ginges:2003qt,Pospelov:2005pr,Engel:2013lsa}. Recent breakthrough sensitivity to $CP$ violation connected to electron spin (that we will refer to as ``paramagnetic EDMs")~\cite{Andreev:2018ayy} established a new limit on the linear combination of the electron EDM $d_e$ and semileptonic nucleon-electron $\bar NN \bar e i\gamma_5 e$
operators, commonly parametrized by a $C_S$ coefficient.
Given rapid progress of the last decade, as well as some additional hopes for increased accuracy (see {\em e.g.} \cite{Vutha:2017pej,Vutha:2018tsz,Fleig:2021qbn}) makes one to revisit the Standard Model (SM) sources of $CP$ violation, and the expected size of the paramagnetic EDMs in the SM. 

SM has two sources of $CP$-violation. First source, undetected thus far, corresponds to the non-perturbative effects parametrized by the QCD vacuum angle $\theta$. Recently it has been shown \cite{Flambaum:2019ejc} that
paramagnetic EDMs are dominated by the two-photon exchange mechanism, and the leading chiral behavior of the hadronic part of the diagram is given by the $t$-channel exchange by $\pi^0,\eta$. $CP$ violation due to $\theta$ comes through the $\pi^0(\eta) \bar NN$ coupling. The result, in combination with the experimental bound \cite{Andreev:2018ayy}, sets the independent limit on $|\theta|< 3\times 10^{-8}$, which is still subdominant to the limit provided by $d_n(\theta)$. 

The second source of the SM $CP$-violation is the celebrated Kobayashi-Maskawa (KM) phase $\delta_{\rm KM} $ \cite{Kobayashi:1973fv}, which is now observed to rather good accuracy in a plethora of flavor transitions in $B$ and $K$ mesons. Observations are often matched by rather precise theoretical predictions, starting from \cite{Bigi:1981qs}. The predictions of EDM-like observables induced by $\delta_{\rm KM}$ thus far can be summarized by two adjectives: small and uncertain. 
The suppression comes from the necessity to involve at least two $W$-bosons and multiple loops \cite{Shabalin:1978rs,Khriplovich:1985jr,Pospelov:1991zt} involving all three generations of quarks. As a result, short distance contributions to quark EDMs do not exceed $10^{-33}\,e\,{\rm cm}$ level~\cite{Czarnecki:1997bu}.  At the same time, it is clear 
that long-distance nonperturbative contributions, typically described as a combination of two transitions changing strangeness by one unit, $\Delta S = \pm1$, dominate $d_n$ and nucleon-nucleon forces \cite{Khriplovich:1981ca,Gavela:1981sk,Flambaum:1984fb,Donoghue:1987dd,McKellar:1987tf}. More recent estimate \cite{Seng:2014lea} places $d_n$ in the ballpark of ${\rm few}\times 10^{-32}\,e\,{\rm cm}$ with a wide order-of-magnitude expected range. It is fair to say that magnitudes of $d_n$ and nucleon-nucleon forces (that feeds into the nuclear-spin-dependent atomic EDMs) cannot be accurately predicted at this point. 

What is the size of paramagnetic EDMs induced by $\delta_{\rm KM}$? Recent estimates of $d_e$ \cite{Yamaguchi:2020eub} (dominated again by long-distance effects) converge at the tiniest value of $\sim 6\times 10^{-40}\,e\,{\rm cm}$, presumably with considerable uncertainties corresponding to hadronic modelling of quark loops. This result is subdominant to the $C_S$ estimate due to the two-photon exchange mechanism in combination with $\Delta S = \pm1$ transitions \cite{Pospelov:2013sca}, that corresponds to equivalent $d_e$ of $\sim 10^{-38}\,e\,{\rm cm}$. To introduce useful notations, this is ${\rm EW}^2{\rm EM}^2$ order effect, where EW/EM stands for electroweak/electromagnetic. 

In this {\em Letter} we demonstrate that the dominant contribution to paramagnetic EDMs associated with the KM $CP$-violation is given by the semileptonic $C_S$ induced in ${\rm EW}^3$ order.  It has an unambiguous answer in the flavor-$SU(3)$ chiral limit, and is calculable to $\sim 30\%$ accuracy that can be further improved. Remarkably, the result reaches the level of $\sim 10^{-35}\,e\,{\rm cm}$ in terms of the $d_e$ equivalent, which is three orders of magnitude larger than previously believed~\cite{Pospelov:2013sca}. 

Our starting point is the expression for the {\em equivalent} $d_e$ that follows from atomic/molecular theory, and defines the linear combination of two Wilson coefficients constrained by the most precise paramagnetic EDM measurements performed with ThO molecule: 
\begin{equation}
\label{deeq}
    d_e^{\rm equiv} = d_e +C_S\times 1.5\times 10^{-20}\,e\,{\rm cm},
\end{equation}
where $e$ is the positron charge. Current experimental limit~\cite{Andreev:2018ayy} stands as $|d_e^{\rm equiv}| < 1.1 \times 10^{-29}\,e\, {\rm cm}$. As per convention, $C_S$ is defined with the Fermi constant factored out, and $\gamma_5$ corresponds to the $\frac12\gamma_\mu (1-\gamma_5)$ definition of the left-handed current:
\begin{equation}
    {\cal L}_{eN} = C_S\frac{G_F}{\sqrt{2}}(\bar e i \gamma_5 e) ( \bar pp + \bar nn).
    \label{eq:CS}
\end{equation}
Our goal is to calculate $C_S(\delta_{\rm KM})$.

\textbf{Leading chiral order $C_S$ calculation}\,---\,
Because of the conservation of the electron chirality in the SM, it is clear that $C_S\propto m_e$. This in turn rules out single photon exchange (EM penguin) as origin of $m_e \bar e i\gamma_5 e$, and one would need either a two-photon mechanism \cite{Pospelov:2013sca,Flambaum:2019ejc} or the EW penguin $Z$-boson exchange/$W$-box diagram. The most crucial property of EW penguins is that although they are  formally of the second order in weak interactions, their size is enhanced by the heavy top, so that the result scales as $G_F^2m_t^2$. EW penguins\footnote{As is well known, EW penguins must also include $W$-box diagrams, and we include both.} induce $B_{s,d} \to \mu^+\mu^-$ decays, and dominate the dispersive part of $K_L \to \mu^+\mu^-$ amplitude. Dropping the vector part of the lepton current (as not leading to $m_e \bar e i\gamma_5 e$), and integrating out heavy $W,Z,t$ particles, one can concisely write down the semileptonic operator as 
\begin{eqnarray}
\label{semileptonic}
{\cal L}_{\rm EWP} = {\cal P}_{\rm EW} \times \bar e \gamma_\mu\gamma_5 e\times \bar s \gamma^\mu(1-\gamma_5) d +(h.c.),
\end{eqnarray}
where 
\begin{eqnarray}
\label{EWP}
{\cal P}_{\rm EW} = \frac{G_F}{\sqrt{2}}\times V_{ts}^*V_{td} \times \frac{\alpha_{\rm EM}(m_Z)}{ 4\pi\sin^2\theta_W} I(x_t),
\end{eqnarray}
and the loop function is given by \cite{Inami:1980fz}
\begin{equation}
\label{IL}
    I(x_t) = \frac34\left( \frac{x_t}{x_t-1}\right)^2\log x_t + \frac14 x_t -  \frac34 \frac{x_t}{x_t-1},~x_t = \frac{m_t^2}{m_W^2}.
\end{equation}
These results are well established, and unlike the case of four-quark operators, the subsequent QCD evolution of (\ref{semileptonic}) introduces only small corrections (see {\em e.g.} \cite{Buras:2012ru}). 

\begin{figure}
  \includegraphics[scale=1]{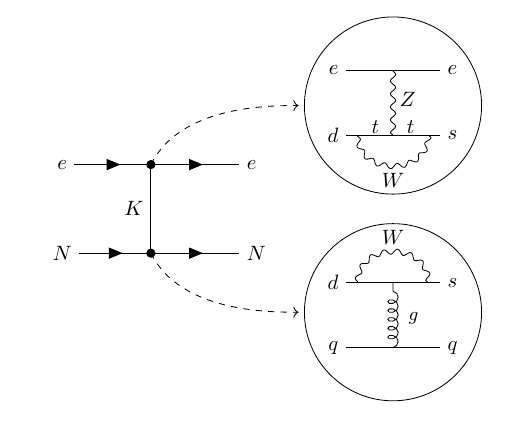}
  \caption{EW$^{3}$ order diagram that dominates in the chiral limit. The top vertex is the $CP$-odd, $P$-even $K_S\bar e i \gamma_5 e $ generated in EW$^{2}$ order, and the bottom vertex is $CP$-even, $P$-odd $K_S\bar NN$ coupling generated at EW$^1$ order. } 
   \label{EW3}
\end{figure}

The most convenient representation of the CKM matrix is when $\delta_{\rm KM}$ enters mostly in $V_{td}$. It enters the imaginary part of ${\cal P}_{\rm EW}$ and couples the axial vector current of leptons to the $\bar s \gamma_\mu(1-\gamma_5) d - \bar d \gamma_\mu(1-\gamma_5) s$ quark current. 
This current can create/annihilate $CP$-even combination of the neutral kaons that (in neglection of small $\epsilon_K$) can be identified with $K_S$ field. Same operator in the muon channel induces $K_S \to \mu^+\mu^-$ meson decay \cite{Isidori:2003ts,DAmbrosio:2017klp}. 
Within chiral perturbation theory,
the axial vector current of leptons is treated as an external left-handed current,
which gives rise to
\begin{align}
	\mathcal{L}_{Uee}
	= 
	-\frac{if_0^2}{2}\mathcal{P}_\mathrm{EW} \times \bar{e}\gamma_\mu \gamma_5e
	\times
	\mathrm{Tr}\left[h^\dagger\left(\partial^\mu U\right) U^\dagger\right]
	+ (h.c.),
	\label{eq:semileptonic_ChPT}
\end{align}
where $U$ is the exponential of the meson octet $M$, $U = \exp[2iMf_0^{-1}]$, in our convention it transforms as $ U'=LUR^\dagger $,
and $h_{ij} = \delta_{i2} \delta_{j3}$.
At linear order, this leads to
$\partial_\mu K \times \bar e \gamma^\mu\gamma_5 e$, 
and upon application of the equation of motion for electrons we arrive to 
\begin{equation}
\label{Kee}
    {\cal L}_{Kee} = -2\sqrt{2}f_0 m_e \bar e i\gamma_5 e \left( K_S \times {\rm Im}{\cal P}_{\rm EW} + K_L \times {\rm Re}{\cal P}_{\rm EW} \right).
\end{equation}
In this expression, $f_0$ is the meson coupling constant, that in the $SU(3)$ symmetric limit is equal to $\simeq 134$\,MeV, and we follow Ref.~\cite{Bijnens:1985kj} conventions. Subsequent $m_s$-dependent corrections renormalize this coupling to $f_0\to f_K\simeq 160$\,MeV. While other $s$-quark 
containing resonances may also contribute, the neutral kaon exchange, Fig. \ref{EW3}, will give the only $m_s^{-1}$-enhanced contribution in the chiral limit. 

We now need to find out how the neutral kaons couple to the nucleon scalar densities, $\bar pp$ and $\bar nn$ that occur due to $\Delta S = \pm 1$ transitions in the EW$^1$ order. Instead of attempting such calculation from first principles (see {\em e.g.} \cite{Shifman:1975tn}) we will use flavor $SU(3)$ relations and connect this coupling to the $s$-wave amplitudes of hadronic decays of strange hyperons, following \cite{Bijnens:1985kj}. It is well known that empirical $\Delta I = 1/2$ rule holds for hyperon decays, and the leading order $SU(3)$ relations fit $s$-wave amplitudes with $O(10\%)$ accuracy. It is strongly suspected that these amplitudes are indeed induced by strong penguins (SP), although this assumption is not crucial for us. With that, one can write down the two types of couplings consistent with $(8_L,1_R)$ transformation properties: 
\begin{equation}
\label{BBM}
{\cal L}_{\rm SP} = -a {\rm Tr}(\bar B \{\xi^\dagger h \xi,B\}) - b {\rm Tr}(\bar B [\xi^\dagger h \xi,B]) +(h.c.).
\end{equation}
In this expression, $B$ is the baryon octet matrix, and 
$\xi = \exp[iMf_0^{-1}]$. Assuming $a$ and $b$ to be real, and taking $f_0=f_\pi$, 
they are fit to be\footnote{
	The overall sign of $a$ and $b$ is not fixed by the hyperon nonleptonic decay
	(the relative sign between $a$ and $b$ is fixed to be negative).
	We use the sign motivated by the vacuum factorization
	of strong penguins~\cite{Shifman:1975tn,Tandean:2002vy}.
	If the overall sign is opposite, it only affects the overall sign of $C_S$ (and $d_e^{\mathrm{equiv}}$)
	and not its absolute value.
}
\begin{equation}
\label{a,b}
a = 0.56 G_Ff_\pi\times [m_{\pi^+}]^2 ;~~b = -1.42 G_Ff_\pi\times [m_{\pi^+}]^2.
\end{equation}
Brackets over $m_{\pi^+}$ indicate that these are numerical values taken, $139.5$\,MeV, rather than $m_u+m_d$-proportional theoretical quantity $m_\pi$.
In the assumption of $a$ and $b$ being real, only the $K_S$ meson couples to nucleons, $2^{1/2}f_0^{-1}((b-a)\bar pp  +2b\bar nn)K_S$, which will provide the dominant contribution. This type of coupling breaks $P$ but respects $CP$ symmetry.   Restoring the CKM factors, one can also include much subdominant coupling to $K_L$ so that we have: 
\begin{eqnarray}
\label{KNN}
    {\cal L}_{KNN} \simeq -\frac{\sqrt{2}\,G_F \times [m_{\pi^+}]^2 f_{\pi}}{|V_{ud}V_{us}|f_0}\times 2.84(0.7\bar pp +\bar nn)\\
    \times\left({\rm Re}(V_{ud}^*V_{us})K_S +  {\rm Im}(V_{ud}^*V_{us})K_L\right). \nonumber
\end{eqnarray}

At the last step, we integrate out the $K$ mesons as shown in Fig.\,\ref{EW3}. Adopting it for a nucleus containing $A=Z+N$ nucleons, one arrives to a straightforward prediction for the $\delta_{\rm KM}$-induced size of the electron-nucleon interaction: 
\begin{equation}
\label{CSanswer}
    C_S \simeq {\cal J} \times \frac{N+0.7Z}{A} \times \frac{13 [m_{\pi^+}]^2f_\pi m_e G_F}{m_{K}^2}\times \frac{\alpha_{\rm EM} I(x_t)}{\pi \sin\theta_W^2}, 
\end{equation}
where ${\cal J}$ is the rephasing invariant combination of the CKM angles, 
\begin{equation}
    {\cal J} = {\rm Im}(V_{ts}^*V_{td}V_{ud}^*V_{us})\simeq 3.1\times 10^{-5},
\end{equation}
that carries about $\sim 6\%$ uncertainty. Notice that the $f_0$ factor in the numerator of (\ref{Kee}) cancels against $f_0$ in the denominator of (\ref{KNN}), and this cancellation would persist even one changes $f_0$ for $f_K$.  

The overall scaling of this formula in the chiral limit and at large $x_t$ is 
\begin{equation}
    G_F C_S \propto {\cal J}G_F^3m_t^2m_em_s^{-1}\Lambda_{\rm hadr}^2.
\end{equation}
where $\Lambda_{\rm hadr}$ is a typical 
hadronic energy/momentum scale.
Notice that this is far more singular behavior with $m_q$ of a light quark than that arising in the chiral-loop-induced expressions for $d_n$. Also notice that the $K_S$ exchange dominates for any conventional parametrization of the CKM matrix, and the role of $K_L$ exchange is to add small pieces of the amplitude that take ${\rm Re}(V_{ud}V_{us}^*){\rm Im}(V_{ts}V_{td}^*)$, arising from $K_S$ exchange, to full  ${\cal J}$. Substituting all SM parameters, we obtain the following leading order result:
\begin{equation}
  C_S({\rm LO}) \simeq 5\times 10^{-16}.
\end{equation}

In order to estimate accuracy of the LO $\sim O(m_s^{-1})$ result, one could try to evaluate the NLO corrections in the expansion over small $m_s$. These corrections can be divided into two groups: A. corrections to the $K\bar NN$ vertex at $m_s\log m_s$ order, B. diagrams that do not reduce to the $t$-channel $K$-meson exchange. Type A corrections involve essentially same diagrams as those appearing in the corresponding corrections to the $s$-wave hyperon decays \cite{Bijnens:1985kj,Jenkins:1991bt}. The analysis of Ref. \cite{Jenkins:1991bt} showed that when the loop corrections are included with the tree-level $a$ and $b$ parameters and the total theoretical result is fit to experimental data, one notices that the tree level values for $a$ and $b$ come out smaller than in (\ref{a,b}), while {\em total} result is rather close to the tree-level fit for $a,b$. This comes mostly from the renormalization of the meson and baryon wave functions. The lesson from this is that the corrections of type A for $KNN$ weak coupling are expected to mirror results of Ref. \cite{Jenkins:1991bt} for $s$-wave amplitudes, and therefore would not deviate substantially from Eq. (\ref{KNN}). 

\begin{figure}[t]
	\centering
 	\includegraphics[width=0.65\linewidth]{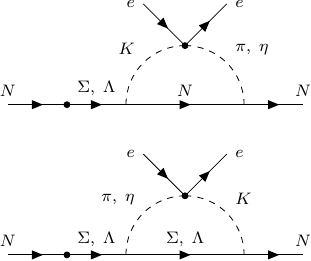}
	\caption{The baryon pole diagrams that contribute to $C_S$ at the NLO level
	in the chiral limit.
	The left vertex is the nucleon-hyperon mixing induced by Eq.~\eqref{BBM},
	while the top vertex is induced by Eq.~\eqref{eq:semileptonic_ChPT}.
	The vertices without black dots are the strong interaction
	with the coupling constants $D$ and $F$.
	The diagrams with the nulceon-hyperon mixing on the right side give the same amount of contribution.}
	\label{fig:NLO}
\end{figure}

We then estimate type B corrections. It turns out that they parametrically dominate over other types of corrections, as the baryon pole diagrams, Fig.~\ref{fig:NLO}, contribute. The $m_s$ scaling of these corrections is set by the ratio of the loop integral, proportional to $m_K$ (at $m_K^2 \gg m_\pi^2$ limit), divided by mass splitting $\Delta m_B$ in the baryon octet, {\em e.g.} $m_\Lambda-m_n$. This quantity scales as $m_s^{-1/2}$ and therefore these baryon pole diagrams dominate the NLO contributions in the chiral limit. They are fully calculable ({\em i.e.} do not depend on unknown counterterms), and the result for these corrections are:
\begin{eqnarray}
\frac{C_{S,NLO}(p)}{C_{S,LO}(p)} 
=\frac{m_K^3(0.77D^2+2.7DF-2.3F^2)}{24\pi f_0^2(m_{\Sigma^+}-m_p)}~~~\\
\frac{C_{S,NLO}(n)}{C_{S,LO}(n)} 
= \frac{m_K^3}{24\pi f_0^2}\left( \frac{(a/b+3)}{2\sqrt{6} (m_\Lambda -m_n)} ~~~\right.\\
\times (-0.44 D^2 + 3.2 DF + 1.3 F^2)\nonumber\\
+\frac{a/b-1}{2\sqrt{2}(m_{\Sigma^0} -m_n)}
\left.(-0.53 D^2 - 1.9 DF + 1.6 F^2)\right)\nonumber.
\end{eqnarray} 
It has been obtained using heavy baryon chiral perturbation theory, 
and $D,F$ are the coupling 
constants characterizing the strength of the $SU(3)$-invariant baryon-meson strong interaction, with $F=0.46,\, D =0.8$ typically used \cite{Bijnens:1985kj}. Since the dominant contribution comes from loops with $K-\pi$ transition, it is appropriate to take $f_0^2 \simeq f_\pi f_K$. Using these numbers, we discover that NLO corrections interfere constructively with LO, and give 30\% correction for the proton, 
and 40\% for the neutron, correspondingly. Combining LO and NLO, we arrive at our final result,
\begin{eqnarray}
\label{LO+NLO}
 C_S({\rm LO+NLO}) \simeq 6.9\times 10^{-16} \nonumber\\
 ~\Longrightarrow~ d_e^{\rm equiv} \simeq 1.0\times 10^{-35}\,e\,{\rm cm}.
\end{eqnarray}
The size of the NLO corrections also allows us to estimate the accuracy of this computation as $O(30\%)$.

As stated in the introduction, this result is much larger than previously believed, and exceeds any contributions of $d_e$ into $d_e^{\rm equiv}$ by at least four orders of magnitude. The enhancement of $C_S$ at EW$^3$ order compared to EW$^2$EM$^2$ can be roughly ascribed to $\alpha_{W}/\alpha_{\rm EM}^2\sim O(10^3)$.
We note that, although translating $C_S$ to $d_e^{\mathrm{equiv}}$ depends on atoms/molecules
that one considers (ThO above),
this dependence is mild and $d_e^{\mathrm{equiv}}$ is within the same ballpark if we instead consider, \emph{e.g.},
Tl or YbF~\cite{Pospelov:2013sca}.

Stepping away from chiral expansion, 
one can formulate the necessary hadronic matrix element that will be required to generate $C_S$ in combination with the dominant Im${\cal P}_{\rm EW}$ channel of Eq.\,(\ref{semileptonic}). The corresponding $d$-to-$s$ transitions need to be taken in the first order, EW$^1$, that break $P$ and $C$ separately but conserve $CP$. 
\begin{eqnarray}
\label{FF}
    \langle N| i(\bar s \gamma_\mu(1-\gamma_5) d - \bar d \gamma_\mu(1-\gamma_5) s) |N\rangle_{\rm EW^1}\\
    \nonumber = \frac{f_S}{m_N} iq_\mu \bar NN + \frac{f_T}{m_N} q_\nu \bar N\sigma_{\mu\nu}\gamma_5N.
\end{eqnarray}
In this formula, $q_\mu$ stands for the momentum transfer. 
It turns out that there are only two form factors on the r.h.s.~of this expression that have the same $CP$ properties as the left-hand side. Moreover, $f_T$ in combination with Im${\cal P}_{\rm EW}$ leads to $CP$-odd $P$-even interactions that do not induce EDMs. Therefore only $f_S$ form factor (that sometimes is called induced scalar) at $q^2\to 0$ is relevant.
We have provided first two terms in  the chiral expansion of $f_S$ for neutrons and protons, so that effectively $f_S \propto a(b) \times m_Nm_K^{-2}+...$.  While we use chiral perturbation theory, in principle, calculation of $f_S$ can be attempted using lattice QCD methods.

Finally, we note that other semi-leptonic operators such as $\bar ee \bar N i \gamma_5 N $ that lead to nuclear-spin-dependent effects are not generated the same way at EW$^3$ order and therefore will be suppressed compared to (\ref{CSanswer}).

\textbf{Conclusions}\,---\,
We have shown that $\delta_{\rm KM}$ induces the $CP$-odd electron nucleon interaction at the level much larger than previous estimates \cite{Pospelov:2013sca}. The main mechanism is not a two-photon exchange, EW$^2$EM$^2$, between electron and the nucleus, but the combination of weak non-leptonic EW$^1$ transition with the semileptonic EW$^2$ electroweak penguin. Although the result is still small, it is not unthinkable that the progress in sensitivity to paramagnetic EDMs may reach the level of $d_e^{\rm equiv}$ in the future.
Indeed, some novel proposals \cite{Vutha:2017pej} envision that statistical sensitivity to paramagnetic EDMs can be brought down to $d_e \sim O(10^{-35}\mathchar`-\mathchar`-10^{-37})\,e\,\mathrm{cm}$.

It is not surprising that the $C_S$ operator can be predicted, at least in the $SU(3)$ chiral expansion, rather precisely. This puts in clear distinction with 
$d_n(\delta_{\rm KM})$ estimate that carries an order of magnitude uncertainty with unclear prospects for improvement. In contrast, the only significant source of uncertainty in $C_S$ is in the induced scalar  form factor (\ref{FF}) that can be improved in the future with the use of lattice QCD methods. 

Even if one takes chiral $SU(3)$ expansion sceptically, it is clear that unique $m_s^{-1}$ (LO) and $m_s^{-1/2}$ (NLO) contributions to $C_S$ identified in our work would not be cancelled - unless completely accidentally - by other contributions, mirroring a similar argument of \cite{Crewther:1979pi} made for $d_n(\theta)$. Therefore, $10^{-35}\,e\,{\rm cm}$ should be adopted as the robust $\delta_{\rm KM}$-induced SM benchmark value for all experiments attempting the search of $d_e$ using electron spins in heavy atoms and molecules. It also allows for establishing the {\em maximum} sensitivity to CP-violating New Physics via $d_e$. Taking a one-loop perturbative scaling, $d_e\propto (\alpha/\pi)m_e\Lambda_{\rm NP}^{-2}$, and equating it to $d_e^{\rm equiv}(\delta_{\rm KM})$ one arrives at the maximum scale that is possible to probe with paramagnetic EDMs: $\Lambda_{\rm NP}^{\rm max} \sim 5\times 10^7$\,GeV.

\vspace{3.5mm}
\begin{acknowledgments}
\textbf{Acknowledgments}\,---\,
Y.E. and M.P. are supported in part by U.S. Department of Energy Grant No.
desc0011842. The Feynman diagrams in this paper are drawn with \texttt{TikZ-Feynman}~\cite{Ellis:2016jkw}.
\end{acknowledgments}

\bibliographystyle{utphys}
\bibliography{CSSM}

\clearpage
\appendix
\onecolumngrid

\renewcommand{\thesection}{S\arabic{section}}
\renewcommand{\theequation}{S\arabic{equation}}
\renewcommand{\thefigure}{S\arabic{figure}}
\renewcommand{\thetable}{S\arabic{table}}
\renewcommand{\thepage}{S\arabic{page}}
\setcounter{equation}{0}
\setcounter{figure}{0}
\setcounter{table}{0}
\setcounter{page}{1}

\end{document}